\documentclass[journal]{IEEEtran}

\usepackage{amsmath}
\usepackage{balance}

\usepackage[colorlinks,urlcolor=blue,linkcolor=blue,citecolor=blue]{hyperref}

\usepackage{comment}
\usepackage[export]{adjustbox}
\usepackage{ifthen}
\usepackage{amssymb}
\usepackage{xcolor}
\usepackage{comment}

\newboolean{showcomments}
\setboolean{showcomments}{true}

\makeatletter
\newcommand{\mynote}[3]{%
  \ifthenelse{\boolean{showcomments}}{%
   \fbox{\bfseries\sffamily\scriptsize#1}%
   {\small$\blacktriangleright$\textsf{\emph{\color{#3}{#2}}}$\blacktriangleleft$}}%
  {%
   \@bsphack
   \@esphack
  }%
}
\makeatother

\definecolor{asparagus}{rgb}{0.53, 0.66, 0.42}

\newcommand{\so}[1]{\mynote{SO}{#1}{red}}

\usepackage{multirow}
\usepackage{multicol}
\usepackage{amsmath}
\usepackage{algorithm}
\usepackage[noend]{algpseudocode}
\usepackage[caption=false,font=footnotesize,labelfont=sf,textfont=sf]{subfig}

\usepackage[export]{adjustbox}

\ifCLASSINFOpdf
\else
\fi
  \usepackage[caption=false,font=footnotesize]{subfig}
\begin{document}
%
\title{FPIRM: Floating-point Processing\\ in Racetrack Memories}
%
%
%

\author{S{\'{e}}bastien Ollivier and
               Xinyi Zhang and
               Yue Tang and
               Chayanika Choudhuri and
               Jingtong Hu and
               Alex K. Jones
\thanks{S. Ollivier, X. Zhang, Y. Tang, C. Choudhuri, J. Hu, and A. K. Jones are with the Department
of Electrical and Computer Engineering, University of Pittsburgh, Pittsburgh,
PA, 15261 USA e-mail: \{sbo15,xinyizhang,yut51,roc74,jthu,akjones\}@pitt.edu}
}

\maketitle

\begin{abstract}
Convolutional neural networks (CNN) have become a ubiquitous algorithm with growing applications in mobile and edge settings. We describe a compute-in-memory (CIM) technique called FPIRM using Racetrack Memory (RM) to accelerate CNNs for edge systems. Using transverse read, a technique that can determine the number of ‘1’s multiple adjacent domains, FPIRM can efficiently implement multi-operand bulk-bitwise and addition computations, and two-operand multiplication. We discuss how FPIRM can implement both variable precision integer and floating point arithmetic. This allows both CNN inference and on-device training without expensive data movement to the cloud. Based on these functions we demonstrate implementation of several CNNs with back propagation using RM CIM and compare these to state-of-the-art implementations of CIM inference and training in Field-Programmable Gate Arrays. During training FPIRM improves by 2$\times$ the efficiency, by reducing the energy consumption by at least 27$\%$ and increasing the throughput by at least 18$\%$ against FPGA.

\end{abstract}

\section{Introduction}
\IEEEPARstart{E}{dge} computing has become increasingly attractive for accelerating machine learning algorithms, such as convolutional neural networks (CNNs), to support needs of mobile applications.  However, edge systems must adhere to constraints often referred to as SWaP (Size, Weight, and Power).  For CNN acceleration, Field Programmable Gate Arrays (FPGA) 
are studied as best possible acceleration engines for low latency small batch inference acceleration while meeting energy requirements of these edge systems.

Spintronic \textit{Racetrack Memory}~\cite{Parkin-08-Science} (RM) is attractive for Compute in Memory (CIM) as it has the necessary density, \textit{i.e.,} between 1-4F$^2$ per cell, while not suffering from  endurance concerns of other tiered memory candidates such as phase-change and resistive memories.  It also has a low energy consumption of circa 0.1pJ~\cite{yu2014energy} per write and a low access latency of circa 1ns generating excitement for use as main memory~\cite{DWM-Main-Memory1,ShiftsReduce}, particularly for SWaP constrained (\textit{e.g.,} edge) systems.

We present \textit{Floating-point Processing in Racetrack Memories} (FPRIM)
.  Core to FPIRM are CIM implementations of multi-operand bulk-bitwise operations and addition operations as well as two-operand multiplication operations.  
We demonstrate floating-point multiplication and addition with FPIRM that enables on-device training using back propagation.  

With FPIRM CIM acceleration of deep learning we achieve as much as 5$\times$ higher performance than state of the art DRAM CIM which leverages ternary (bulk-bitwise and summation) weight calculations~\cite{dracc,elp2im} with a nearly 50\% reduction in power.  FPIRM is 2.8$\times$ faster and more than 3$\times$ more energy efficient for integer precision (multiplication and addition) than the state-of-the-art RM CIM.
We also achieve 18--74\% performance improvement and 26--81\% reduction in power compared to a low-energy FPGA for 32-bit floating-point precision online training targeting small to moderate CNNs. 
In particular, FPIRM makes the following contributions:
\begin{itemize}
    \item FPIRM is, to our knowledge, the first RM CIM  approach to implement floating-point addition and multiplication.
    \item We propose floating-point CIM designed to conduct multi-operand floating-point addition.
    \item We show that FPIRM outperforms and provides better efficiency for both CIM (inference) and FPGA (training) targeting edge systems.
\end{itemize}

The remainder of this paper is organized as follows: First, we provide the necessary background on CNNs and RM. Next, we describe the basic concepts of
FPIRM, starting with its architecture and how to perform integer operations. We then explain how to perform floating-point multiplication followed by a step-by-step explanation on how to perform FP multi-operand addition. Then, our experimental results compare the improvements of FPIRM  with state of the art FPGA architecture, for three ML training benchmarks. Finally, we reach conclusions.

\section{Background}

In this section, we first introduce the elements that compose the CNN inference, then we introduce the additional operations required for training and the difference between these algorithms. In a second time, we introduce the fundamental of RM, and its advantages as a CIM for CNN.

\subsection{Convolutional Neural Network}
\newcommand{\myMatrix}[1]{\mathbf{{#1}}}
CNNs are primarily based on the convolution operation, which is a windowed point-wise multiplication accumulation of multiple channels of input features with a set of weights to generate output features
. As an example, for the input features $\myMatrix{I}$ and weights $\myMatrix{K}$ of size $N\times R_{in}\times C_{in}$ and $M\times N\times 3\times 3$, respectively, the convolution operation for the window at $m$ (output channel index), $r$ (row), $c$ (column) is:

\footnotesize \[
Conv(\myMatrix{I},\myMatrix{K})(m,r,c)= \sum_{n=0}^{N-1}\sum_{j=0}^{2} \sum_{t=0}^{2}  \myMatrix{K}_{m,n,j,t}\times\myMatrix{I}_{n,r+j,c+t}
\vspace{-.05in}
\] \normalsize\\
where $M$ is the number of output channels, $N$ is the number of input channels, $R_{in}\times C_{in}$ is the size of an input feature map.
The inference operation requires convolution steps broken up with activation layers 
composed of \textit{pooling layers} to reduce dimensionality of input matrices through average or maximum value operations and \textit{ReLU function}, a linear function that will output the input if positive and zero otherwise. Once these convolution layers are completed,  \textit{fully-connected layers} are used to provide the classification result. The fully-connected layers can be mathematically written as $ReLU(\myMatrix{W}\myMatrix{x}+\myMatrix{b})$.

Training of the CNN includes forward-propagation, loss back-propagation, and weight update. 
During the forward-propagation, which is same as in inference, the values at each activation layer are stored for the weight update. 
The loss is calculated by a loss function such as Cross-entropy loss~\cite{de2005tutorial}.  
%
%
After calculating the loss of the last layer, the loss is propagated layer by layer until reaching the first layer of CNN model, by a process that includes
weight rotation,  convolution, and channel-wise accumulation.
Based on the loss back propagation, the weights are updated in each layer individually based on the weight, activation, and the loss of the activation is determined, typically using gradient descent. 
The operations in weight updates are depth-wise convolution, element-wise multiplication, and element-wise subtraction.

While deep learning with CNNs presumes calculations with floating-point values, CNN inference calculations can often be reduced to integer computation with as few as 8-bits achieving reasonable accuracy.  Recent DRAM CIM work has shown that in many cases this can be further reduced to ternary $w\in\{-1,0,1\}$~\cite{dracc} or even binary $w\in\{0,1\}$ computations~\cite{sim2018nid} operations to replace the multiplications.  However, online training for all but the simplest CNNs still requires full 32-bit floating-point computations to work properly.  Without this accuracy, the weight updates can be ineffective and possibly even detrimental.

In the next section we describe basic of Racetrack memory that serves as the foundation for FPIRM CIM to accelerate these CNN functions.




\subsection{Racetrack Memory Fundamentals}


\begin{figure}[tbp]
\centering
\includegraphics[width=\columnwidth]{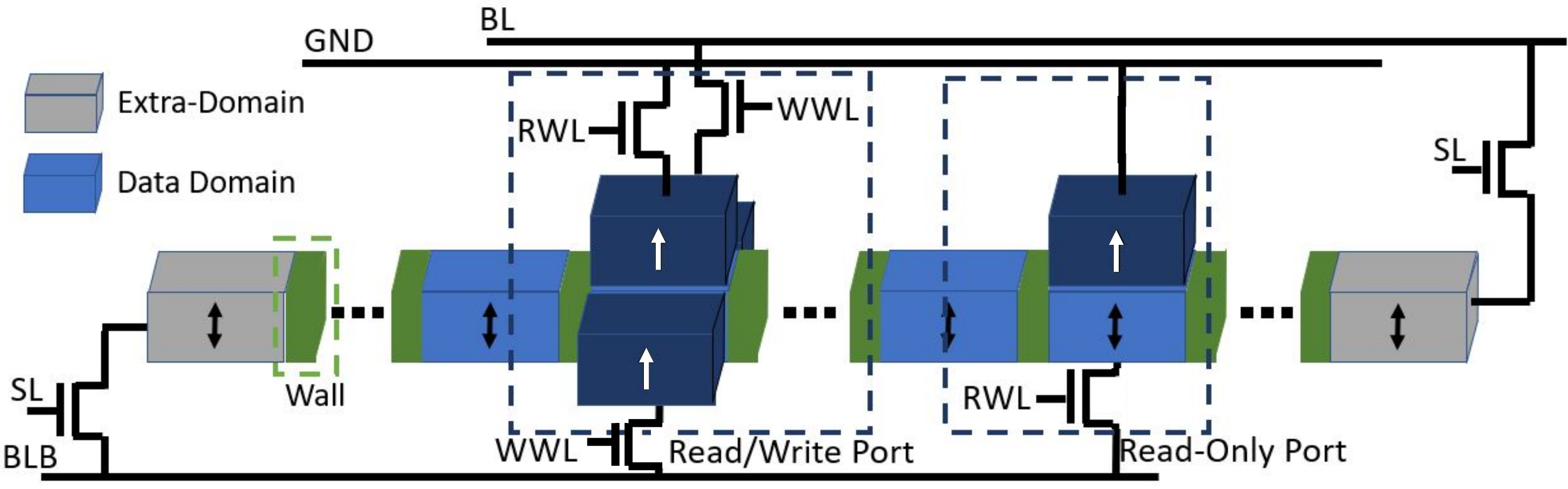}
\caption{Anatomy of a domain-wall memory nanowire. }
\label{DWMzoom}
\label{fig:nanowire}
\vspace{-.2in}
\end{figure}

Spintronic RM is made of ferromagnetic nanowires consisting of many magnetic domains separated by domain walls (DWs) as shown in Fig.~\ref{DWMzoom}. Each domain has its own magnetization direction 
such that binary values are represented by the magnetization direction of each domain, either parallel/antiparallel to a fixed reference. For a planar nanowire, several domains share one/few access point(s) (APs) for read and write operations~\cite{zhang2012perpendicular}. DW motion is controlled by applying a short current pulse laterally along the nanowire governed by \texttt{SL}. Random access requires {\em shifting} the target domain to {\em align} it with an AP (dark blue) and apply a current to {\em read} or {\em write} the target bit.  To avoid data loss when shifting, the blue domains store actual data while the grey domains are overhead domains to prevent data loss. 
Shift-based writing (Read/Write Port)~\cite{DWM_Tapestri} allows slower current writes to be replaced with orthogonal shifts from fixed magnetic alignment domains to reduce latency and energy.

RM, like many other novel memories including resistive memory CIM crossbars~\cite{chi2016prime}, has also received significant attention for CIM, particularly for deep learning~\cite{yu2014energy,CNN_DWM,PIRM}.
In the next section we describe our FPIRM technique to build CIM approach with RM that can operate at multiple levels of previsions from binary/ternary weight inference to full floating-point precision required for CNN training.

\section{FPIRM}

\begin{figure*}[tbp]
\centering
\subfloat[Subarray built from tiles and DBCs\label{subfig:subarray}]{
    \includegraphics[height=1.6in]{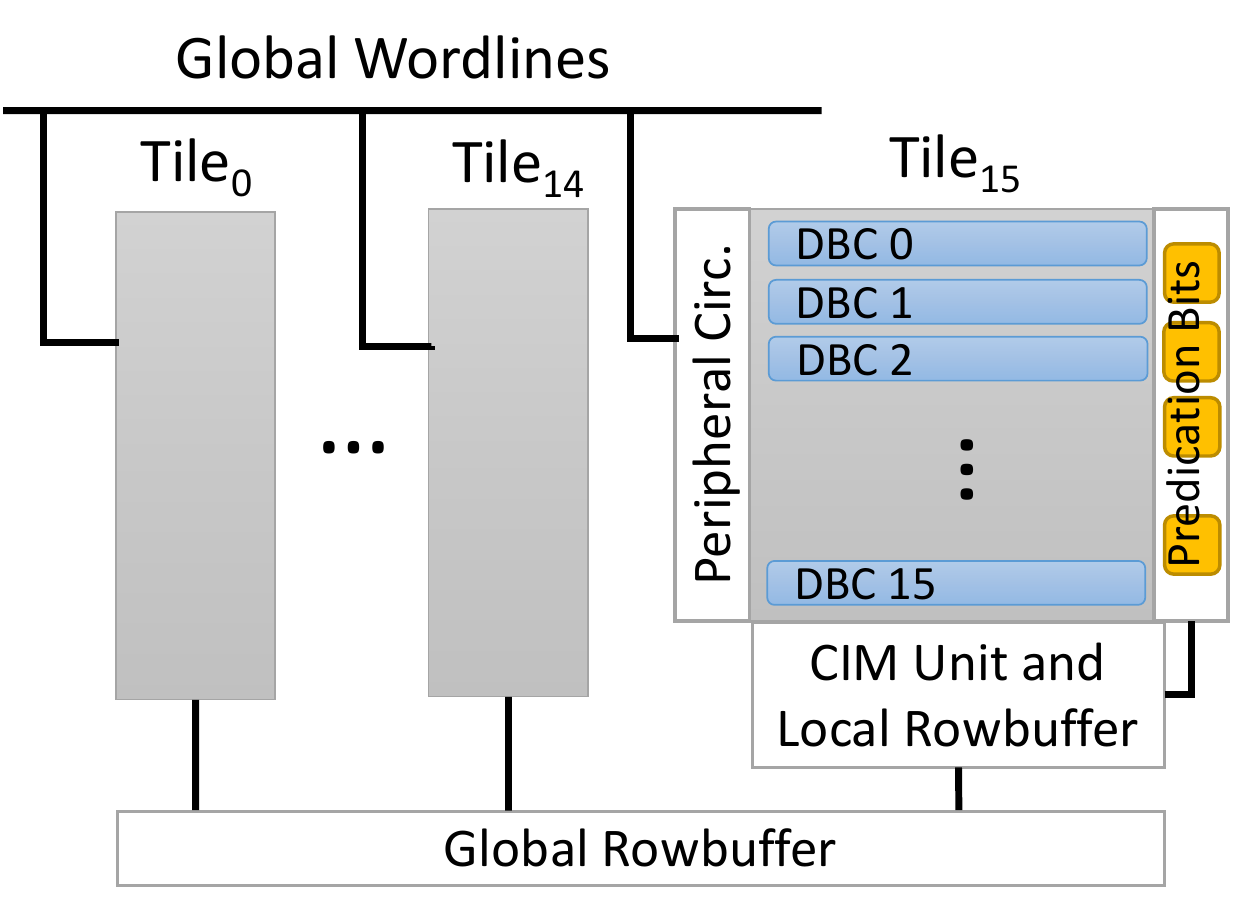}
}\hspace{-.1in}
\subfloat[DBC design: 2 APs for TR~~\label{subfig:DBC}]{
    \includegraphics[height=1.6in]{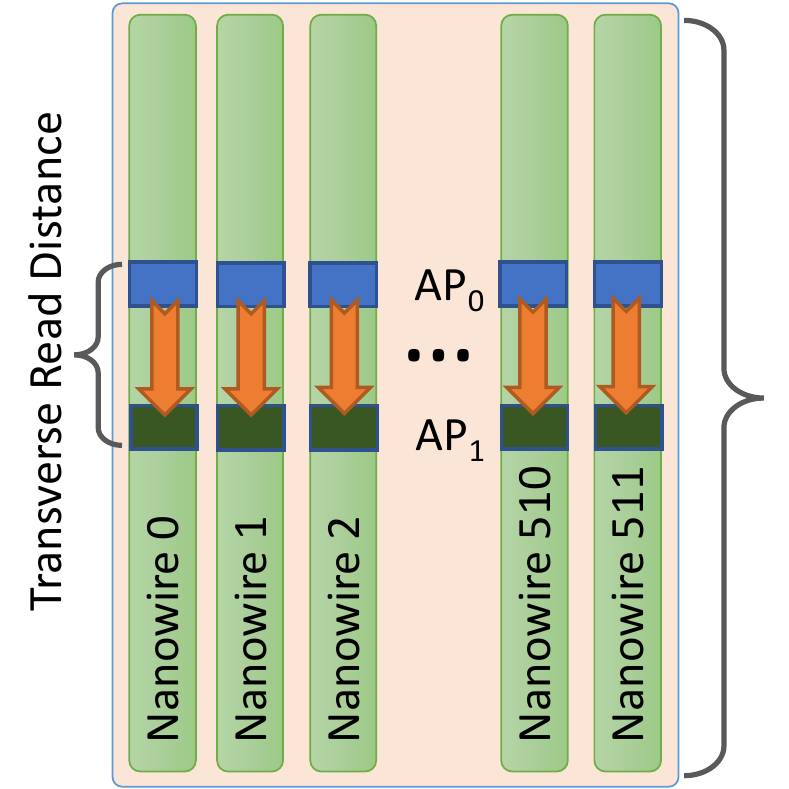}
}\hspace{-.2in}
\subfloat[CIM unit for logic, arithmetic, and shifting\label{subfig:PIMUnit}]{
    \includegraphics[height=1.6in]{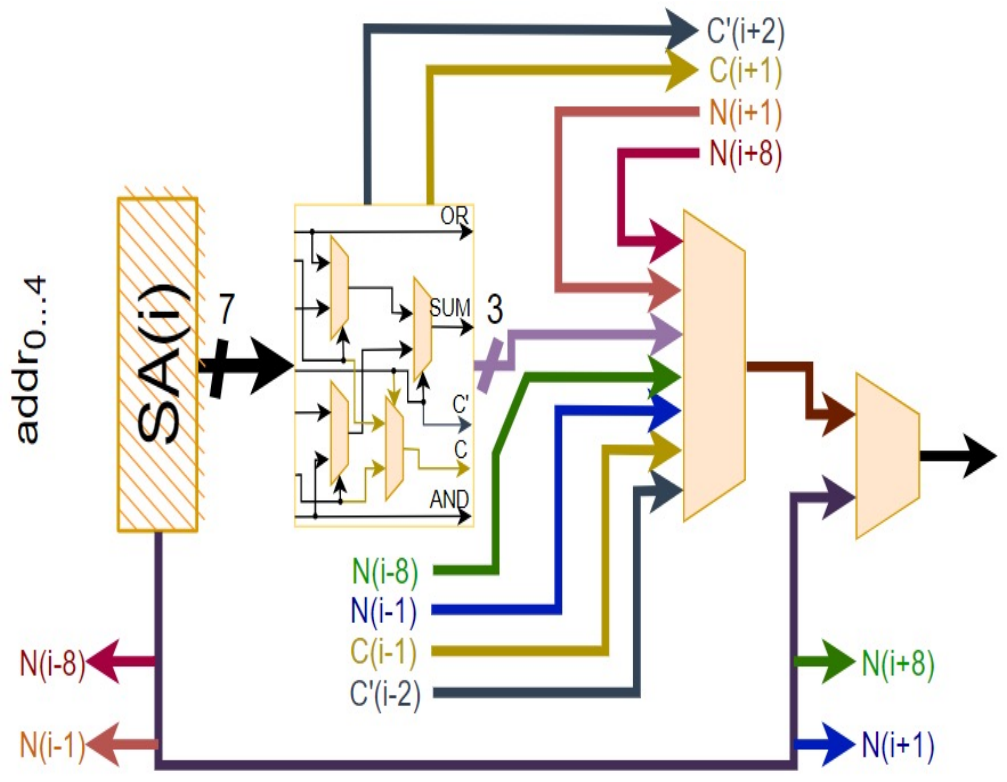}
}\\
\caption{Racetrack memory architecture following rank, bank, subarray, and tile conventions.  All tiles are decomposed into DBCs with some tiles (\textit{e.g.,} one per subarray) augmented for transverse access and CIM.}
\label{fig:architecture}
\vspace{-.2in}
\end{figure*}

The memory architecture concept behind FPIRM is shown in Fig.~\ref{fig:architecture}.  We follow a DRAM-inspired hierarchical organization consisting of ranks and banks constructed from subarrays built with tiles (Fig.~\ref{subfig:subarray}).  Each tile is constructed from bundles of RM nanowires shifted together and referred to as a domain-block cluster (DBC)~\cite{TapeCache,ShiftsReduce}. A DBC can accommodate $D$ rows with parallel access of all bits belonging to the same row through the parallel APs (Fig.~\ref{subfig:DBC}).  $D \in\{16,32,64\}$ is the number of data domains per nanowire.  Each tile maintains a $512 \times 512$ shape, akin to DRAM.

To enable CIM processing, a tile may be extended in two ways.  A second AP is added to the DBCs in that tile to allow a current to traverse all the domains between the two APs indicated by the orange arrow (Fig.~\ref{subfig:DBC}).  If spaced within a prescribed \textit{transverse read distance} (TRD), Transverse Read (TR) can distinguish between resistance levels based on the number of `1's between the APs much like a multi-level cell~\cite{roxy2020novel}.  Using TR, the local rowbuffer is extended with a \textit{CIM Unit} that retains a fast (bypass) path for a standard read, but can also convert the one counts from a TR into multi-operand logic and arithmetic based on the TRD (shown for $\text{TRD}=7$ in Fig.~\ref{subfig:PIMUnit}).

Multi-operand \texttt{AND} and \texttt{OR} are naturally determined by sensing the highest (all ones) or lowest (all zeros) resistance levels.  Operations of fewer operands can be accomplished by padding with ones or zeros as appropriate.  Unlike prior processing using memory approaches FPIRM includes logic to directly compute \texttt{XOR} from the 1's count which is also the sum $S$ for addition.  All of these \textit{bulk-bitwise} operations may be computed in parallel across the entire memory row.  We also can compute a carry $C$ and \textit{super carry} $C'$ with minimal additional logic.  Addition of $\text{TRD}-2$ operands may be computed directly by activating each nanowire in sequence.  For nanowire $N_i$,  $S_i$ is written to AP$_0$ of $N_i$ ($S_i \rightarrow N_{i,0}$), while similarly $C_i \rightarrow N_{i+1,1}$ and $C'_i \rightarrow N_{i+2,0}$ in parallel enabling a carry chain using navy and yellow connections in the CIM.  

The CIM block allows \textit{logical} left and right shifts by both one (orange and blue) and eight (red and green) positions.  These \textit{logical} shifts are different from RM nanowire shifting which aligns different domains with access points (up and down).  The shift by one position along with a small number of predication bits~\cite{9065566} support multiplication using partial product addition
.  Furthermore, by computing the $S$, $C$, and $C'$ bits in parallel seven operands can be reduced to three, allowing multiplication and reduction over addition to be computed in $O(n)$ time where $n$ is the operand width~\cite{PIRM}. 

The process to compute integer (or fixed-point) computation using FPIRM is shown in Algorithm~\ref{alg:int-operations}.  Addition of $\text{TRD}-2$ operands may be computed directly activating each nanowire in sequence.  For nanowire $N_{i}$, $i$ represents the nanowire that corresponds to a particular bit position of data in a memory row (see Fig.~\ref{subfig:DBC}).  $S_i$, computed using the TR of $N_i$ and the CIM Unit (Fig.~\ref{subfig:PIMUnit}) is written to AP$_0$ of $N_i$ ($S_i \rightarrow N_{i,0}$) where $N_{i,j}$ refers to $N_i$ at AP $j$.  Similarly $C_i \rightarrow N_{i+1,1}$ and $C'_i \rightarrow N_{i+2,0}$.  

The \textsc{Add}$(O[1..5],w)$\footnote{$l\neq0$ can be specified to allow an offset, but we explain for $l=0$.} function of Algorithm~\ref{alg:int-operations} describes this in more detail.  For each nanowire in sequence, starting with the least significant bit, the $S_i$ is computed using the \texttt{XOR} and written to $O[0]_i$, which corresponds to $N_{i,0}$.  Given we have $ones_i$ from the SAs such that $\sum_{k=0}^6 O[k] > i \implies ones_i = 1$ otherwise $ones_i=0$ we can compute the result using multiplexers such that $M_0$ selects $ones_0$ or $ones_2$ based on $ones_1$, $M_1$ selects $ones_4$ or $ones_6$  based on $ones_5$ and $S$ selects $M_0$ or $M_1$ based on $ones_3$.  Similarly, $C$ selects $ones_1$ or $ones_5$ based on $ones_3$  and writes the value in $O[6]_{i+1}$, or $N_{i+1,1}$.  $C'$ is $ones_3$ which is written to $O[0]_{i+2}$ or $N_{i+2,0}$.  By $i=2$ all values of $O[0..6]_{i+2}$ may contain data.  As the most significant bit is reached, values are not written above $N_{w-1}$.  Moreover, just as $S,C,C'$ can be written to $N_i$, $N_{i+1}$, and $N_{i+2}$, simultaneously, when $k$ values are packed into a row, we can activate $k$ nanowires such that $\forall \{0..k-1\} N_{i+k*w}$ generates and writes $S,C,C'$ in parallel.

If it is necessary to add more than $\text{TRD}-2$ numbers, we use \textsc{CSA-Reduction}$(O[0..6])$.  \textsc{CSA-Reduction} is akin to a Carry-Save Adder (CSA) with more than 3 inputs.  A CSA adder computes the $S$ and $C_{OUT}$ from three inputs (\textit{e.g.,} $A,B,C_{IN}$) in parallel before doing a traditional add of $S$ and $C_{OUT}$, which requires traversing the carry chain only once.  Similarly \textsc{CSA-Reduction} computes $\forall i$ $S_i$ in bulk-bitwise fashion, followed by $C_i \rightarrow N_{i+1}$, concluding with $C'_i \rightarrow N_{i+2}$.  While \textsc{Add} takes $w$ cycles to sum five operands, \textsc{CSA-Reduction} takes a constant time to reduce seven operands to three operands.  Only once there are five or fewer remaining operands is \textsc{Add} called.  

We see this use in \textsc{Multiply}$(O_A,O_B,w)$.  First, each bit of $O_A$ is used to determine whether a shifted copy of $O_B$ is copied as a partial product of the multiplication.  We leverage the connections from $N_i \rightarrow \{N_{i-1},N_{i+1}\}$ in Fig.~\ref{subfig:PIMUnit} to accomplish this along with the predication registers~\cite{9065566} shown in Fig.~\ref{subfig:subarray}.  Operand $OP_A$ is logically shifted left while $OP_B$ is logically shifted left so that the predication register which causes a shifted copy of $OP_B$ to be retained can be pulled from the zeroth bit of the rowbuffer to simplify the logic that selects the predication bit source.  In this instance the least significant bit of each packed operand must have a connection to the predication register.   After preparing the partial products they are reduced using \textsc{CSA-Reduction} until five or fewer values remain.  These values are summed using an \textsc{Add}.

\algdef{SE}[VARIABLES]{Variables}{EndVariables}
    {\algorithmicvariables}
    {\algorithmicend\ \algorithmicvariables}
\algnewcommand{\algorithmicvariables}{\textbf{Global variables}}

\definecolor{forestgreen}{RGB}{34, 139, 34}
\algnewcommand{\LeftComment}[1]{\(\triangleright\)#1} 
\algrenewcommand{\algorithmiccomment}[1]{{\color{forestgreen}\(\triangleright\)#1}}

\algrenewcommand\algorithmicindent{1.0em}%

\begin{algorithm}[tbp]
\footnotesize
\caption{Integer/Fixed-point Multiply and Add~\cite{PIRM}}
\label{alg:int-operations}
\begin{algorithmic}[1]

\State \Comment{Fixed-point multiply}
\Function{Multiply}{$O_A,O_B,w$}

\For{$i \gets 0:w-1$} \Comment{w-bit operands}
    \State $P[i] \gets 0$ \Comment{Initialize Partial Product}
    \State \Comment{Predicated copy on $O_A[i]$ of $O_B << i$}
    \State $O_A[0]$ ? $P[i] \gets $\textsc{Copy} $O_B$ 
    \State $O_A \gets O_A >> 1$ \Comment{align $i+1$ with 0 position}
    \State $O_B \gets O_B << 1$ \Comment{prepare $M_B << i$} 
\EndFor
$t \gets \lceil w/7 \rceil; r \gets w\mod7 $
\While {$(t > 1) || (r>5)$}
\For{$i \gets 0:t-1$} 
    \State $P[3i...3i+2]\gets$ \textsc{CSA-Reduction}($P[7i..7i+6]$)
    \If {$r>3$} 
      \State $P[3t...3t\!+\!2]\!\!\gets$\textsc{CSA-Reduction}($P[7t..7t\!+\!r]$)
     \Else 
       \State $P[3t...3t+r] \gets $ $P[7t..7t+r]$)
     \EndIf
     \EndFor
$t \gets  \lceil (t*3+r) / 7 \rceil; r \gets (t*3+r) \mod 7$
\EndWhile
    \State $P[0] \gets $ \textsc{Add}($P[0..r]$,w*2)
    \State \Return{P[0]}
\EndFunction

\State

\State \Comment{Reduce 7 operands of bit-width $w$ to a Sum ($S$), Carry ($C$), and Super Carry ($C'$)}
\Function{CSA-Reduction}{$O[0...6]$}
    \State \Comment{Intrinsic CIM operations allowed using TR~\cite{roxy2020novel}}
    \State \Comment{All bits $i$ in width $w$ in parallel~\cite{PIRM}}
    \ForAll{$i$ in $w$}
        \State $S_i \gets O[0]_i \oplus O[1]_i \oplus ... \oplus O[6]_i$
        \State $ones_i \gets \sum_{k=0}^6 O[k]_i$
        \State $C_i \gets 2 \leq  ones_i < 4 \; \text{\texttt{OR}} \; ones_i \; \geq 6$
        \State $C'_i \gets ones_i \geq 4$
    \EndFor
    \State \Return{$S,C,C'$}
\EndFunction

\State

\State \Comment{Add 5 operands of bit-width $w$ starting at lower-bound bit $l$ (default $=0$), using parallel carry and super carry chains~\cite{PIRM}, and return the Sum ($S$)}
\Function{Add}{$O[1..5],w,l=0$}
    \State \Comment{All bits $i$ in sequence to form carry chain}
    \State \Comment{$O[0],O[6]$ reserved for $C',C$; $S$ overwrites $O[0]$}
    \State $u = l + w$
    \For{$i \gets l:u-3$}
        \State $O[0]_i \gets O[0]_i \oplus O[1]_i \oplus ... \oplus O[6]_i$
        \State $ones_i \gets \sum_{k=0}^6 O[k]_i$
        \State $O[6]_{i+1} \gets 2 \leq  ones_i < 4 \; \text{\texttt{OR}} \; ones_i \; \geq 6$
        \State $O[0]_{i+2} \gets ones_i \geq 4$
    \EndFor
    \State \Comment{Compute $S,C$ for penultimate bit, $S$ for final bit}
    \State $O[0]_{u-2} \gets O[0]_{u-2} \oplus O[1]_{u-2} \oplus ... \oplus O[6]_{u-2}$
    \State $ones_{u-2} \gets \sum_{k=0}^6 O[k]_{u-2}$
    \State $O[6]_{u-1} \gets 2 \leq  ones_{u-2} < 4 \; \text{\texttt{OR}} \; ones_{u-2} \; \geq 6$
    \State $O[0]_{u-1} \gets O[0]_{u-1} \oplus O[1]_{u-1} \oplus ... \oplus O[6]_{u-1}$
    \State \Return{$O[0]$} \Comment{$O[0]$ is the final sum}
\EndFunction
\end{algorithmic}
\end{algorithm}

These operands alone form the basis to conduct binary, ternary, and integer/fixed point CNN inference.  In the next sections, we describe floating-point multiplication and reduction over addition required for accurate back propagation used for CNN online training.



\section{Floating-point Two-Operands Multiplication}
\label{Sec:FPMult}

To conduct floating-point multiplication the operands must be separated into their main components of sign, exponent, and mantissa via masking using bulk-bitwise operation.  Using a combination of integer arithmetic on the extracted elements we can compute floating-point multiplication.  Given the nature of CNN training where point-wise products are reduced over addition, we leave the products separated to support multi-operand summation.

Presuming the 32-bit floating-point standard of a 23-bit mantissa, an 8-bit exponent (biased by $2^8-1$), packed into 64-bits to provide space for the multiplication.  First the mantissa $M_i$, exponent $E_i$, and sign $S_i$ of each operand $i\in\{A,B\}$ is masked off with an \texttt{AND} operation and stored separately.  The implied leading `1' is also restored to each mantissa with an \texttt{OR} operation.  
%
Using integer operations we multiply the two 24-bit mantissas allowing expansion to 48-bits.  Because $1.0 \leq \{M_A,M_B\} < 2.0$, then $1.0 \leq M=M_A\times M_B < 4$.  To normalize we use the top bit, which if `1' executes a predicated normalization right shift enabled by the CIM unit.  We then add the exponents $E = E_A + E_B + -127 (+1)$ using multi-operand integer arithmetic, adding negative 127 due to the normalization factor, and adding 1 if normalization was required when generating $M$.  
%
Finally, the sign bit $S = S_A \, \text{\texttt{XOR}} \, S_B$ to determine the resulting sign of the multiplication.  

We follow the flow in Algorithm~\ref{alg:fp-mult}.  First the mantissas of both operands are extracted and reconstructed to include the implicit `1' using bulk bitwise operations as shown in lines~\ref{line:mantissa-mask}--\ref{line:mantissa-mask-end}.  Calling \textsc{Multiply} from Algorithm~\ref{alg:int-operations} the product mantissa is calculated.  The normalization process checks the 48th bit (line~\ref{line:normalization}) and uses this to populate the predication register for a predicated shift operation required to normalize mantissas $M \geq 2.0$ (line~\ref{line:pred-normalize}).  This requires that the predication register must be able to select between bit 0 for multiplication or 47 from the row-buffer to populate the predication register.  

To continue we mask off the exponents (lines~\ref{line:exp-mask}--\ref{line:exp-mask-end}).  Next the new exponent must be calculated, we sum the two extracted exponents $E_A, E_B$ but must subtract (add negative) 127, which is loaded into the memory location indicated by $offset$, to deal with the exponent normalization factor.  If the mantissa was shifted for normalization, we increase the exponent by one through a predicated storage of 0x800000 into an otherwise empty location denoted as $one$.  Executing the \textsc{Add} function sums the exponents $E_A,E_B$, 0x7F800000 (-127), and optionally 0x800000 as a normalization if the mantissa required normalization.  With a 5th 0x0 operand, the new exponent is calculated by adding $w=8$-bits starting at bit $l=23$ (line~\ref{line:exponent-add}).  Note, if $E_{31}$ is `1' the exponent has gone out of range (overflow) because either $E > 255$ or $E < 0$.  In lines~\ref{line:sign}--\ref{line:sign-end} the sign is similarly extracted and computed using an \texttt{XOR}.  Finally in $M, E,$ and $S$ are returned, but the mantissa remains using 47-bits.

\begin{algorithm}[tbp]
\footnotesize
\caption{Floating-point Multiplication}
\label{alg:fp-mult}
\begin{algorithmic}[1]
\State \algorithmicvariables: $OP_A, OP_B, M, E, S$
\State  \Comment {Inputs: $OP_{A,B}$ Operands}
\State  \Comment {Outputs: $M$ = Mantissa, $E$ = Exponent, $S$ = Sign}
\State
\Function{FPMultiply}{$OP_A,OP_B$}
\State \Comment {Mask off and multiply the mantissas}
\label{line:mantissa-mask}
\State $M_A \gets (OP_A$ \texttt{AND} 0x7FFFFF$)$ \texttt{OR} 0x800000
\State $M_B \gets (OP_B$ \texttt{AND} 0x7FFFFF$)$ \texttt{OR} 0x800000
\label{line:mantissa-mask-end}
\State $M \gets$ \textsc{Multiply} $(M_A,M_B)$
\State \Comment{normalize if $M \geq 2.0$}
\State $norm \gets (M$ \texttt{AND} 0x800000000000$)$ 
\label{line:normalization}
\State \Comment{Predicated $M$ normalization shift testing $norm$}
\State $norm$ ? $M \gets M >> 1$ 
\label{line:pred-normalize}
\State \Comment{Mask off, add the exponents with normalization} 
\label{line:exp-mask}
\State $E_A \gets (OP_A$ \texttt{AND} 0x7F800000$)$
\State $E_B \gets (OP_B$ \texttt{AND} 0x7F800000$)$
\label{line:exp-mask-end}
\State $offset \gets$ 0xC0800000 \Comment{~-127 to correct exponent}
\State $one \gets$ 0x0
\State $norm$ ? $one \gets $ 0x800000 \Comment{increase exponent if $M\!\geq\!2.0$}
\State \Comment{Add exponents, note overflow if $E_{31} = 1$}
\State $E \gets $ \textsc{Add}$(E_A,E_B,offset,one,$0x0,8,23)
\label{line:exponent-add}
\State \Comment{Mask off and determine final sign}
\label{line:sign}
\State $S_A \gets (OP_A$ \texttt{AND} 0x80000000$)$
\State $S_B \gets (OP_B$ \texttt{AND} 0x80000000$)$
\State $S \gets (S_A$ \texttt{XOR} $S_B)$
\label{line:sign-end}
\State \Return {$M,E,S$}
\EndFunction
\end{algorithmic}
\end{algorithm}

We leave $M$, $E$, and $S$, decomposed to facilitate reduction over floating-point addition described in the next section.


\section{Floating-Point Multi-Operands Addition}
\label{Sec:FPAdd}



The addition required in the CNN application is a reduction over addition to combine products of pointwise multiplication within a convolution window.  Given the sign, mantissa, and exponents of these products are already segregated, the first step is to determine the maximum exponent within the convolution window and then normalize all the remaining exponents and their mantissas to this value.  This process is similar to determining the maximum value during pooling.  Copies of the values, in this case the exponents, are sequentially tested via TR from most to least significant bit.  If the TR$\geq$1, each word with a `1' is re-written shifted left by one position.  The other words are overwritten with a zero vector.  This is accomplished by reading the value in the shifted position enabled by the CIM unit.  A predicated row reset based on the current tested bit of the current value is used to zero the ``eliminated'' values.  If the TR$=$0 all values are cycled through and written back in shifted form.  The shifting permits the predication value to be referenced from the same location in the rowbuffer.  At the end a TR is conducted and the maximum value is read and written shifted back by 8 as permitted by the CIM unit.  This process is repeated with all the participants of the summation over reduction to obtain the maximum value.  

For mantissa normalization, we subtract the local exponent with the maximum and use the difference to normalize the mantissa via shifting.  
First the lowest difference bit is read into the predication register and a predicated logical right shift (read and shift using the CIM unit) is issued.  The next bit is for a logical shift by two and we continue by powers of two for each subsequent bit. 
%
The connection to $N_{i+8}$, $N_{i+1}$,
are used to accelerate this procedure.  

Next the sign bit of each operand is used as a predication value to invert the mantissa and store `1' in the neighboring row for 2's complement inversion and summed using integer addition.  The resulting summation is subsequently normalized to the 23 bits by complementing negative numbers, normalizing the exponent to correct value.  This conducted by creating a copy of the mantissa that is logically shifted to find the first instance of `1' which when found triggers adding the appropriate exponent offset and activates logical shifts to the original mantissa.  

We conduct floating-point addition following the flow in Algorithm~\ref{alg:fp-add}.  The function begins by searching for the maximum exponent among the operands in groups of TRD$=7$ seven (lines~\ref{line:find-max-exponent}--\ref{line:find-max-exponent-end}) using \textsc{FindMax} from the helper functions in Algorithm~\ref{alg:fp-add-helper} as follows: Starting with the most significant bit ($p$) an \texttt{OR} operation ($isAnyOne$) detects any `1's at that position (line~\ref{line:test-ones}).  If there is at least one `1' then we eliminate values with `0's at that position because they will be smaller by resetting the rowbuffer and writing them back as 0x0.  The algorithm does this by reading each exponent in turn from AP$_0$, inverts the exp with an \texttt{XOR} with 1 ($nIsOne$) then \texttt{AND}s this with $isAnyOne$ (lines~\ref{line:test-local-one}--\ref{line:test-local-one-end}).  A predication bit $pred$ is extracted from the row buffer at position $p=31$ and controls the rowbuffer reset (line~\ref{line:reset-RB}).  The nanowires are shifted toward AP$_0$ and the value is written back to AP$_1$ (lines~\ref{line:round-robin}--\ref{line:round-robin-end}, see Fig.~\ref{subfig:DBC}) such that after a round of TRD ($=7$ shown in the algorithm) each exponent is written back and stored in its correct original position.  If there are no `1's in this position, the predicate logic is never true because $isAnyOne$ is zero (line~\ref{line:test-ones}), so none of the exponents are eliminated (line~\ref{line:reset-RB}).   

Note, each exponent is logically shifted left by one each time, this is to facilitate reading the predicate from bit $p$ each round.  This ensures the predicate is only selected from position 0, 31, and 47.  The final result is logically shifted right by eight to put the maximum exponent value in the correct alignment (line~\ref{line:shift-right-8}).  The main function uses \textsc{FindMax} in $\log_{\text{TRD}=7}$ fashion to find the overall maximum exponent now stored in E.

\begin{algorithm}[tbp]
\footnotesize
\caption{Floating-point Addition}
\label{alg:fp-add}
\begin{algorithmic}[1]
\State \algorithmicvariables: $M[0..n-1], E[0..n-1], S[0..n-1]$
\State  \Comment {Inputs: $M[0..n-1], E[0..n-1], S[0..n-1]$ Operands}
\State  \Comment {Outputs: $M$ = Mantissa, $E$ = Exponent, $S$ = Sign}
\Statex
\State \Comment{Conduct Floating-Point Addition of $n$ values}
\Function{FPAdd}{M[0..n-1], E[0..n-1], S[0..n-1]}
\State \Comment {Find maximum exponent searching groups of TRD (7)}
\label{line:find-max-exponent}
\State $t \gets \lceil n/7 \rceil; r \gets n\mod7 $
\While {$t>1$}
\For{$i \gets 0:t-1$} 
   \State E[i] $\gets$ \textsc{FindMax}(E[7i..7i+6])
\EndFor
\For{$i \gets 0:r-1$}
    \State E[t+i] $\gets$ E[7*t+i]
\EndFor
\State $t \gets  \lceil (t+r) / 7 \rceil; r \gets (t+r) \mod 7$
\EndWhile
\For{$i \gets r:6$}
    \State E[$i$] $\gets 0$
\EndFor
\State E = \textsc{FindMax}{E[0..6]} 
\label{line:find-max-exponent-end}
\State \Comment {Normalize Mantissas and convert to 2's complement}
\label{line:norm-2s}
\For{$i \gets 0:n-1$}
  \State $m[2*i] = $ \textsc{NormMantissa}$($M[$i$],E,E[$i$]$)$
  \State $m[2*i+1] = $ 0x0
  \label{line:invert-on-sign}
  \State S[$i$] ? $m[2*i] \gets m[2*i]$ \texttt{XOR} 0xFFFFFFFFFFFFFFFF
  \State S[$i$] ? $m[2*i+1] \gets $0x1
\EndFor
\label{line:invert-on-sign-end}
\label{line:norm-2s-end}
\State \Comment {Sum Mantissas}
\label{line:mantissa-reduction}
\State $t \gets \lceil n*2/7 \rceil; r \gets 2*n\mod7 $
\While {$(t > 1) || (r>5)$}
\For{$i \gets 0:t-1$} 
    \State $m[3i...3i+2]\gets$ \textsc{CSA-Reduction}($m[7i..7i+6]$)
    \If {$r>3$} 
      \State $m[3t...3t\!+\!2]\!\!\gets$\textsc{CSA-Reduction}($m[7t..7t\!+\!r]$)
     \Else 
       \State $m[3t...3t+r] \gets $ $m[7t..7t+r]$)
     \EndIf
     \EndFor
\State $t \gets  \lceil (t*3+r) / 7 \rceil; r \gets (t*3+r) \mod 7$
\EndWhile
\label{line:mantissa-reduction-end}
    \State $M \gets $ \textsc{Add}($m[0..r]$,64)
\State $Sum \gets$ \textsc{NormSum}$(M,E)$
\State \Return{E[0]}
\EndFunction
\end{algorithmic}
\end{algorithm}

Next all of the mantissas must be normalized to the maximum exponent and after normalization if their sign is negative inverted in twos complement for summation (Algorithm~\ref{alg:fp-add}, lines~\ref{line:norm-2s}--\ref{line:norm-2s-end}).  The first step uses the \textsc{NormMantissa} helper function (Algorithm~\ref{alg:fp-add-helper}).  First the exponent is subtracted from the maximum exponent (lines~\ref{line:sub-exp}--\ref{line:sub-exp-end}), then each bit is inspected to normalize the mantissa.  Again the predication bit is extracted from the same position $p$ as in the \textsc{FindMax} function.  If the highest two bits are true, the mantissa is shifted by 128 or 64 places which results in setting it to 0x0 (lines~\ref{line:out-of-scope}--\ref{line:out-of-scope-end}).  For the next three bits (5:3) 4, 2, and 1 predicated shifts by 8 are executed (line~\ref{line:pred-shift-8}), followed by 4, 2, and 1 predicted shifts by 1 (line~\ref{line:pred-shift-1}).  Note, these mantissas are still ``normalized'' to bit position 47 not 23 from the multiplication in a prior step, so a shift by 32 is still in scope.  
After normalization, two storage locations for each value are allocated.  If the value is positive the first location gets the mantissa and the second get 0x0.  Using predication from the sign bit (also bit position 31) the mantissa is inverted and the second location is written 0x1 (Algorithm~\ref{alg:fp-add} lines~\ref{line:invert-on-sign}--\ref{line:invert-on-sign-end}).  

\begin{algorithm}[tbp]
\footnotesize
\caption{Addition Helper Functions}
\label{alg:fp-add-helper}
\begin{algorithmic}[1]

\Statex
\State \Comment{Find the Maximum Among 7 values}
\Function{FindMax}{E[0..6],w=8,o=23}
\State $p \gets o + w$
\For{$i \gets o : o + w - 1$}
\State $isAnyOne \gets $ \texttt{OR} E[0..6]  \Comment {TR$\geq1$}
\label{line:test-ones}
     \For{$j \gets 0 : 6$}
        \State $nIsOne \gets $E[j]$_o$ \texttt{XOR} 0xFFFFFFFFFFFFFFFF
        \label{line:test-local-one}
        \State $pred \gets isAnyOne$ \texttt{AND} $nIsOne$
        \label{line:test-local-one-end}
        \State $RB \gets $E[j]$ << 1$

        \State $pred_p $ ? RESET $RB$ \Comment{test at bit $p \implies RB \gets 0$}
        \label{line:reset-RB}
\State SHIFT rows upward \Comment{Shift DWM DBC, reindex E}
\label{line:round-robin}
        \State E[6] $\gets RB$
        \label{line:round-robin-end}
    \EndFor
\EndFor
\State \Return{TR E[0..6] $>> 8$}
\label{line:shift-right-8}
\EndFunction

\Statex
\State \Comment{Normalize a Mantissa based on Exponent Difference}
\Function{NormMantissa}{M, Max, E,w=8,o=23}
\State $p \gets o + w$
\State $t \gets $ E \texttt{XOR} 0xFF800000 \Comment{Invert Exponent}
\label{line:sub-exp}
\State $S \gets $\textsc{Add}$($Max,$t$,0x800000,0,0,w+1,o$)$
\label{line:sub-exp-end}
\For {$i \gets 7:6$}
\label{line:out-of-scope}
  \State $isOne \gets S_p$; $S \gets S << 1$
  \State $isOne$ ? M $\gets$ 0x0
\EndFor
\label{line:out-of-scope-end}
\For {$i \gets 5:0$}
   \State $k \gets i \mod 3$
   \State $isOne \gets S_p$; $S \gets S << 1$
   \For{$j \gets 0:2^k$}
    \If{$i<3$} $isOne$ ? M $\gets$ M $>> 1$
    \label{line:pred-shift-1}
    \Else $\;isOne$ ? M $\gets$ M $>> 8$
    \label{line:pred-shift-8}
    \EndIf
  \EndFor
\EndFor
\State \Return{M}
\EndFunction

\Statex
\State \Comment{Normalize Mantissa based on Summation and prepare Sum}
\Function{NormSum}{M, E}
\State \Comment {Get sign from MSB and if negative, get 2's Complement}\label{line:mantissa-magnitude}
\State $sign \gets $ M
\For{$i \gets 0:3$} \Comment {Move the sign to bit 31}
\State $sign >> 8$
\EndFor
\State $sign_{31}$ ? M $\gets$ M \texttt{XOR} 0xFFFFFFFFFFFFFFFF
\State $sign_{31}$ ? M $\gets$ \textsc{Add}$($M,0x1,0,0,0,64$)$
\label{line:mantissa-magnitude-end}
\State \Comment{Copy mantissa, $<<$ to find first `1', norm exp with}
\State \Comment{predicate, shift mantissa to bit 23}
\State $m \gets $M; $m << 1$; $seenOne \gets $ 0x0
\State \Comment{If Mantissa is higher than bit 47, increase the exp}
\State \Comment{Shift the mantissa right to bit position 47}
\For{$i \gets 15:0$}
  \State $seenThisOne \gets m$
  \texttt{OR} $seenOne$ \label{line:seen-one}
  \State $seenOneFirst \gets seenThisOne$ \texttt{XOR} $seenOne$ \label{line:seen-one-end}
  
  \State $seenOneFirst_{63}$ ? $expAdd \gets i << 23$ 
  \label{line:add-offset}
  \State $seenOne \gets seenThisOne$
  \State $seenOne_{63}$ ? $M \gets M >> 1$ \label{line:before-47}
  \State $m \gets m << 1$
\EndFor
\State \Comment{If Mantissa is lower than bit 47, decrease the exp}
\For{$i \gets 1:47$}
  \State $seenThisOne \gets m$ \texttt{OR} $seenOne$
  \State $seenOneFirst \gets seenThisOne$ \texttt{XOR} $seenOne$
  \State $seenOneFirst_{63}$ ? $expAdd \gets -i << 23$ 
  \State $seenOne \gets seenThisOne$
  \State $nSeenOne\!\gets\!seenOne $ \texttt{XOR} 0xFFFFFFFFFFFFFFFF
    \State \Comment{From 47..24, shift the mantissa right to bit position 23}
  \If{$i < 23$}
  \State $seenOne_{63}$ ? $M \gets M >> 1$ \label{line:before-23}
  \EndIf
    \State \Comment{From 22...0, shift the mantissa left to bit position 23}
  \If{$i > 23$}
  \State $nSeenOne_{63}$ ? $M \gets M << 1$ \label{line:after-23}
  \EndIf
  \State $m \gets m << 1$
\EndFor
\State $exp \gets $ \textsc{ADD}$($E,$expAdd$,0,0,0,8,23$)$ \Comment {Add offset to exp} \label{line:the-end}
\State $sum \gets m$ \texttt{AND} 0x7FFFFF \Comment{Strip leading `1'}
\State $sum \gets sum$ \texttt{OR} $exp$ \texttt{OR} $sign$ \Comment{Recombine}
\State \Return $sum$ \label{line:the-end-end}
\EndFunction
\end{algorithmic}
\end{algorithm}

Once normalized and converted to twos complement form, the mantissas can be summed.  Similar to multiply, \textsc{CSA-Reduction} is used to reduce operands from $7 \rightarrow 3$ until there are $\text{TRD}-2=5$ or fewer which are then summed using \textsc{Add} (lines~\ref{line:mantissa-reduction}--\ref{line:mantissa-reduction-end}).  The last step is to again normalize the resulting mantissa and to reassemble the floating point number.  Normalization is done using \textsc{NormSum} helper function (Algorithm~\ref{alg:fp-add-helper}).  Since this addition used twos complement logic we extract the sign from most significant bit of the full 64-bit cell.  This becomes the predicate to invert and add 0x1 to make the mantissa positive (lines~\ref{line:mantissa-magnitude}--\ref{line:mantissa-magnitude-end}).  We then look for the first `1' in the mantissa in relationship to bit 47.  

The first bit position requires two storage locations, one to store whether a `1' has been seen $seenOne$ and a second $seenThisOne$ which looks for a `1' at this bit position.  $seenThisOne$ is true if we have seen a `1' previously or there is a one in this round (line~\ref{line:seen-one}.  The predicate $seenOneFirst$ comes from \texttt{XOR} which is only true on the first `1' (line~\ref{line:seen-one-end}) and the $expAdd$ value is only set once (line~\ref{line:add-offset}).  Once $seenOne$ is set, we start shifting M to be aligned with bit 23, which requires right shifts if found before bit 23, shown with predicated shifts on $seenOne$ for bits 62:48 (line~\ref{line:before-47}) and 47:24 (line~\ref{line:before-23}).  If $seenOne$ is still not seen by bit 22, we start shifting left governed by the $seenOne$ complement $nSeenOne$ until the `1' is found (line~\ref{line:after-23}) .    The remainder of the function is to combine the normalization exponent offset $expADD$ with E, strip bit 23 and combine the sign, exponent, and mantissa per the floating point standard (lines~\ref{line:the-end}--\ref{line:the-end-end}).  

Note, these algorithms are designed to show the feasibility of the function.  In some cases, optimizations for system performance or code optimizations for expediency may have been excluded to maintain clarity.  For example, while shown here, we can complete the decomposition and recomposition only at the beginning or ending of the full benchmark when communicating with a host processor.  Additionally, while shown for 63-bits, we can use the lower 32 bits from the $sign$ shift first and then work from the $m$ to only pull predicates from position 31 (selecting from three positions, 47, 31, 0).  

The control for these algorithms comes from the hosts/memory controller.  Control of \textbf{for}, \textbf{if}, \textbf{while}, etc. control constructs are governed by the host as they are deterministic and can be entirely unrolled.  Moreover, these can be distributed via single instruction multiple data (SIMD) execution throughout the system (e.g., via different subarrays) for massive parallelism.  Only the predicated instructions use data-based control, and these presume the instructions will be executed or a nop in its place to remain in lock step with the SIMD execution.  Finally, while shown for 32/64 bits, given the row size is 512, 8 items can be packed per row and computed in parallel.

\section{Additional Operations for Back Propagation}

During back propagation weight matrices must be rotated 180 degrees, which is equivalent to swapping the values of these relatively small (3$\times$3 up to 11$\times$11) along vertical and horizontal bisecting line of the matrix.  We use FPIRM PIM to mask off the individual values of each row using \texttt{AND}, logically shift to the correct position, and recombine using \text{OR}.  Additionally, the weight update operation: $W'=W-L_R\times\Delta W$ where the new weight $W'$ is a function of the previous weight $W$ the learning weight $L_R$ and the weight difference $\Delta W$ calculated via gradient descent method.  We also use floating-point FPIRM CIM to compute this function.

\section{Results}

FPIRM enables multiple precision modes from binary weight used for inference to floating-point required for effective training. Thus we compare FPIRM for inference against state-of-the-art DRAM CIM using ternary weights~\cite{dracc,elp2im} and RM using integer weights~\cite{CNN_DWM} as well as FPIRM for training using floating-point operations against energy-efficient FPGAs suitable for Edge systems: Xilinx ZU19EG (Lenet-10)~\cite{liu2017fpga} and ZCU102 (Alexnet and VGG-16)~\cite{yue2022}.
The energy and latency parameters of accessing RM and TR in FPIRM are provided by~\cite{yu2014energy,roxy2020novel}. 
The latency and energy consumption for the CIM unit architecture extensions in Fig.~\ref{fig:architecture} were determined by implementing the design with the Cadence ASIC Flow targeting 45nm technology.

\setlength\tabcolsep{2pt}
%
%

\setlength\tabcolsep{2pt}
\begin{table}[t]
\centering
\caption{FPIRM compared to accelerators}
\label{tab:Result}
\begin{tabular}{l|c|c|c|c}
\hline
\multicolumn{5}{c}{\textbf{Inference Improvement Compared to CIM}}\\\hline
\textbf{Benchmark} & \textbf{Target} &  \textbf{Throughput}  & \textbf{Power} & \textbf{Efficiency} \\
 & & FPS & W & FPS/W\\
\hline
\textbf{Lenet-5} & DRAM~\cite{elp2im} & 8330 & -- & --\\
Ternary~\cite{dracc} & FPIRM & 32075 & 0.028 & 1.1$\cdot10^6$ \\\hline
\multicolumn{2}{l|}{FPIRM Improvement} & 3.85$\times$ & $\O$ & $\O$\\
\hline
\textbf{Alexnet} & DRAM~\cite{dracc} & 84.8 & 2 & 42.4\\
Ternary~\cite{dracc} & FPIRM & 490 & 0.93 & 526\\\hline
\multicolumn{2}{l|}{FPIRM Improvement} & 5.78$\times$ & 1.94$\times$ & 12.4$\times$\\
\hline
\textbf{Lenet-5} & RM~\cite{CNN_DWM} & 59 & 0.017 & 13291\\
Integer & FPIRM & 163 & 0.006 & 44169\\\hline
\multicolumn{2}{l|}{FPIRM Improvement} & 2.76$\times$ & 2.33$\times$ & 3.32$\times$\\
\hline
\textbf{Alexnet} & RM~\cite{CNN_DWM} & 32.1 & 5.89 & 5.45\\
Integer & FPIRM & 90.5 & 4.99 & 18.13\\\hline
\multicolumn{2}{l|}{FPIRM Improvement} & 2.81$\times$ & 1.18$\times$ & 3.33$\times$\\
\hline
\multicolumn{5}{c}{\textbf{Training Improvement Compared to FPGA}}\\\hline
\textbf{Benchmark} & \textbf{Target} &  \textbf{Throughput}  & \textbf{Power} & \textbf{Efficiency} \\
 & & GFLOPS & W & GFLOPS/W\\
\hline
\multirow{2}{*}{\textbf{Lenet-10}} 
& FPGA~\cite{liu2017fpga} & 86.12 & 14.23 & 6.05 \\
& FPIRM & 101.5 & 2.76 & 36.77 \\
\hline
\multicolumn{2}{l|}{FPIRM Improvement} & 1.18$\times$ & 5.16$\times$ & 6.08$\times$ \\
\hline
\multirow{2}{*}{\textbf{Alexnet}} 
& FPGA~\cite{yue2022} & 34.52 & 7.74 & 4.46 \\
& FPIRM & 50.72 & 5.65 & 8.97 \\
\hline
\multicolumn{2}{l|}{FPIRM Improvement} & 1.47$\times$ & 1.36$\times$ & 2.01$\times$ \\
\hline
\multirow{2}{*}{\textbf{VGG-16}} 
& FPGA~\cite{yue2022} & 46.99 & 7.71 & 6.09 \\
& FPIRM & 81.95 & 5.7 & 14.37 \\
\hline
\multicolumn{2}{l|}{FPIRM Improvement} & 1.74$\times$ & 1.35$\times$ & 2.36$\times$ \\

\hline

\end{tabular}
\vspace{-.2in}
\end{table}

\subsection{CNN Inference}
\label{SubSec:InferenceResult}


During the CNN inference phase, precision can be tuned based on required accuracy.  Reduced precision can provide a lower-latency result \textit{in situ}, which is particularly valuable for edge networks with small batch sizes.   For instance, integer, ternary, or binary weight calculation reduces the complexity of addition and multiplication to simpler integer functional units while providing sufficient accuracy compared to more expensive floating-point computation.  In fact, ternary and binary forms convert multiplication to \textit{much} simpler bulk-bitwise (\textit{e.g.,} \texttt{XOR}) operations. 

Using bulk-bitwise ternary weight CNN inference FPIRM is more than 3$\times$ faster than state-of-the-art DRAM CIM~\cite{dracc,elp2im} with an approximately 2$\times$ power advantage leading to an order of magnitude efficiency advantage for Alexnet\footnote{Power and energy data was not reported for the Lenet-5 DRAM CIM implementation~\cite{elp2im} and is noted as a ``--'' in Table~\ref{tab:Result}.}.  In fact, ternary weight CNN inference with FPIRM is 2--3$\times$ faster than even simpler binary weight CNN inference using DRAM CIM~\cite{PIRM}.   Using integer operations, FPIRM can outperform by nearly 3$\times$ and provides more than 3$\times$ the efficiency of the latest RM CIM~\cite{CNN_DWM}.
The results are detailed in Table~\ref{tab:Result}. 
In the next section, we present FPIRM result on CNN training.

\subsection{CNN Training}
\label{SubSec:TrainingResult}

As training requires a high accuracy and large datasets, which typically can be most efficiently accelerated with a graphics processing unit (GPU).  However, given \textit{in situ} training for low latency with small batch sizes and to maintain SWaP of edge systems GPUs may not be practical for their relatively high power.  Sending these large datasets to the cloud for GPU acceleration is also impractical.  Given CIM has yet to demonstrate CNN training with floating-point precision, we compare with FPGAs accelerators, which are emerging for \emph{in situ} edge CNN training~\cite{liu2017fpga,yue2022}.  
FPIRM is competitive, even outperforming FPGAs by 18--74\% with a significant improvement in power.  We demonstrate a more than 2$\times$ improvement in efficiency even as the complexity of the CNN increases; FPIRM for Alexnet is 2$\times$ more efficient, while VGG-16 is 2.36$\times$ more efficient. 
Thus, not only is FPIRM demonstrating that CNN training is possible using CIM, it may even be more practical than FPGAs.  When coupled with the high capacity and low energy consumption of RM-based memory, the capabilities for SWaP constrained edge acceleration of deep learning and beyond are impressive and worthy of further exploration.

\section{CONCLUSION}

FPIRM is the first, to our knowledge, approach to enable full CNN architectures in memory, with multiple precision capabilities suitable for tuning both inference and training operations. While floating-point operations have always been a major roadblock for in memory processing, FPIRM can perform these operations efficiently at a speed and energy consumption improving over FPGA technology. In particular, FPIRM is between 18$\%$ and 74$\%$ faster in term of throughput, and at least 26$\%$ better in term of energy, resulting in an efficiency improvement of more than 2$\times$ compared to state-of-art FPGAs for small to moderate sized CNNs.  
FPIRM is the first CIM architecture that is sufficiently re-configurable to provide capabilities and improvements over state-of-the-art techniques for both \textit{in situ} CNN inference and training for edge computing.

\section{ACKNOWLEDGMENT}

This work was supported in part by the NSF under grants CNS-1822085, CNS-2133267, the National Security Agency, and Laboratory of Physical Sciences.  We would like to thank Dr. Xulong Tang and Sheng Li for their consultation on this manuscript.

\bibliography{pim}

\begin{thebibliography}{10}
\providecommand{\url}[1]{#1}
\csname url@samestyle\endcsname
\providecommand{\newblock}{\relax}
\providecommand{\bibinfo}[2]{#2}
\providecommand{\BIBentrySTDinterwordspacing}{\spaceskip=0pt\relax}
\providecommand{\BIBentryALTinterwordstretchfactor}{4}
\providecommand{\BIBentryALTinterwordspacing}{\spaceskip=\fontdimen2\font plus
\BIBentryALTinterwordstretchfactor\fontdimen3\font minus
  \fontdimen4\font\relax}
\providecommand{\BIBforeignlanguage}[2]{{%
\expandafter\ifx\csname l@#1\endcsname\relax
\typeout{** WARNING: IEEEtran.bst: No hyphenation pattern has been}%
\typeout{** loaded for the language `#1'. Using the pattern for}%
\typeout{** the default language instead.}%
\else
\language=\csname l@#1\endcsname
\fi
#2}}
\providecommand{\BIBdecl}{\relax}
\BIBdecl

\bibitem{Parkin-08-Science}
S.~S.~P. Parkin, M.~Hayashi, and L.~Thomas, ``Magnetic domain-wall racetrack
  memory,'' \emph{Science}, vol. 320, no. 5874, pp. 190--194, Apr. 2008.

\bibitem{yu2014energy}
H.~Yu, Y.~Wang, S.~Chen, W.~Fei, C.~Weng, J.~Zhao, and Z.~Wei, ``Energy
  efficient in-memory machine learning for data intensive image-processing by
  non-volatile domain-wall memory,'' in \emph{2014 19th Asia and South Pacific
  Design Automation Conference (ASP-DAC)}.\hskip 1em plus 0.5em minus
  0.4em\relax IEEE, 2014, pp. 191--196.

\bibitem{DWM-Main-Memory1}
\BIBentryALTinterwordspacing
D.~Wang, L.~Ma, M.~Zhang, J.~An, H.~H. Li, and Y.~Chen, ``Shift-optimized
  energy-efficient racetrack-based main memory,'' \emph{Journal of Circuits,
  Systems and Computers}, vol.~27, no.~05, p. 1850081, 2018. [Online].
  Available: \url{https://doi.org/10.1142/S0218126618500810}
\BIBentrySTDinterwordspacing

\bibitem{ShiftsReduce}
\BIBentryALTinterwordspacing
A.~A. Khan, F.~Hameed, R.~Bl\"{a}sing, S.~S.~P. Parkin, and J.~Castrillon,
  ``Shiftsreduce: Minimizing shifts in racetrack memory 4.0,'' \emph{ACM Trans.
  Archit. Code Optim.}, vol.~16, no.~4, dec 2019. [Online]. Available:
  \url{https://doi.org/10.1145/3372489}
\BIBentrySTDinterwordspacing

\bibitem{dracc}
Q.~Deng, L.~Jiang, Y.~Zhang, M.~Zhang, and J.~Yang, ``Dracc: a dram based
  accelerator for accurate cnn inference,'' in \emph{2018 55th ACM/ESDA/IEEE
  Design Automation Conference (DAC)}, 2018, pp. 1--6.

\bibitem{elp2im}
X.~Xin, Y.~Zhang, and J.~Yang, ``Elp2im: Efficient and low power bitwise
  operation processing in dram,'' in \emph{2020 IEEE International Symposium on
  High Performance Computer Architecture (HPCA)}.\hskip 1em plus 0.5em minus
  0.4em\relax IEEE, 2020, pp. 303--314.

\bibitem{de2005tutorial}
P.-T. De~Boer, D.~P. Kroese, S.~Mannor, and R.~Y. Rubinstein, ``A tutorial on
  the cross-entropy method,'' \emph{Annals of operations research}, vol. 134,
  no.~1, pp. 19--67, 2005.

\bibitem{sim2018nid}
J.~Sim, H.~Seol, and L.-S. Kim, ``Nid: processing binary convolutional neural
  network in commodity dram,'' in \emph{2018 IEEE/ACM International Conference
  on Computer-Aided Design (ICCAD)}.\hskip 1em plus 0.5em minus 0.4em\relax
  IEEE, 2018, pp. 1--8.

\bibitem{zhang2012perpendicular}
Y.~Zhang, W.~Zhao, D.~Ravelosona, J.-O. Klein, J.-V. Kim, and C.~Chappert,
  ``Perpendicular-magnetic-anisotropy cofeb racetrack memory,'' \emph{Journal
  of Applied Physics}, vol. 111, no.~9, p. 093925, 2012.

\bibitem{DWM_Tapestri}
R.~Venkatesan, M.~Sharad, K.~Roy, and A.~Raghunathan, ``Dwm-tapestri-an energy
  efficient all-spin cache using domain wall shift based writes,'' in
  \emph{Proc. of DATE}, 2013, pp. 1825--1830.

\bibitem{chi2016prime}
P.~Chi, S.~Li, C.~Xu, T.~Zhang, J.~Zhao, Y.~Liu, Y.~Wang, and Y.~Xie, ``Prime:
  A novel processing-in-memory architecture for neural network computation in
  reram-based main memory,'' \emph{ACM SIGARCH Computer Architecture News},
  vol.~44, no.~3, pp. 27--39, 2016.

\bibitem{CNN_DWM}
B.~Liu, S.~Gu, M.~Chen, W.~Kang, J.~Hu, Q.~Zhuge, and E.~H.-M. Sha, ``An
  efficient racetrack memory-based processing-in-memory architecture for
  convolutional neural networks,'' in \emph{2017 IEEE International Symposium
  on Parallel and Distributed Processing with Applications and 2017 IEEE
  International Conference on Ubiquitous Computing and Communications
  (ISPA/IUCC)}.\hskip 1em plus 0.5em minus 0.4em\relax IEEE, 2017, pp.
  383--390.

\bibitem{PIRM}
\BIBentryALTinterwordspacing
S.~Ollivier, S.~Longofono, P.~Dutta, J.~Hu, S.~Bhanja, and A.~K. Jones,
  ``{PIRM:} processing in racetrack memories,'' arXiv, Tech. Rep. 2108.01202,
  2021. [Online]. Available: \url{https://arxiv.org/abs/2108.01202}
\BIBentrySTDinterwordspacing

\bibitem{TapeCache}
R.~Venkatesan, V.~Kozhikkottu, C.~Augustine, A.~Raychowdhury, K.~Roy, and
  A.~Raghunathan, ``Tapecache: a high density, energy efficient cache based on
  domain wall memory,'' in \emph{Proc. of ISLPED)}, 2012, pp. 185--190.

\bibitem{roxy2020novel}
K.~Roxy, S.~Ollivier, A.~Hoque, S.~Longofono, A.~K. Jones, and S.~Bhanja, ``A
  novel transverse read technique for domain-wall “racetrack” memories,''
  \emph{IEEE Transactions on Nanotechnology}, vol.~19, pp. 648--652, 2020.

\bibitem{9065566}
M.~Lenjani, P.~Gonzalez, E.~Sadredini, S.~Li, Y.~Xie, A.~Akel, S.~Eilert, M.~R.
  Stan, and K.~Skadron, ``Fulcrum: A simplified control and access mechanism
  toward flexible and practical in-situ accelerators,'' in \emph{2020 IEEE
  International Symposium on High Performance Computer Architecture (HPCA)},
  2020, pp. 556--569.

\bibitem{liu2017fpga}
Z.~Liu, Y.~Dou, J.~Jiang, Q.~Wang, and P.~Chow, ``An fpga-based processor for
  training convolutional neural networks,'' in \emph{2017 International
  Conference on Field Programmable Technology (ICFPT)}.\hskip 1em plus 0.5em
  minus 0.4em\relax IEEE, 2017, pp. 207--210.

\bibitem{yue2022}
Y.~Tang, X.~Zhang, P.~Zhou, and J.~Hu, ``Ef-train: Enable efficient on-device
  cnn training on fpgathrough data reshaping for online adaptation or
  personalization,'' \emph{ACM Transactions on Design Automation of Electronic
  Systems (TODAES)}, 2022, accepted, preprint appears at arXiv.org.

\end{thebibliography}
\bibliographystyle{IEEEtran}
\balance

\end{document}